# Ultrafast non-volatile flash memory based on van der Waals heterostructures

Lan Liu[1], Yi Ding[1], Jiayi Li[1], Chunsen Liu[1, 2, *] and Peng Zhou[1, *]

[1]State Key Laboratory of ASIC and System, School of Microelectronics, Fudan University, Shanghai 200433, China

[2]School of Computer Science, Fudan University, Shanghai 200433, China

[*]**Corresponding authors**

Emails: chunsen_liu@fudan.edu.cn

Emails: pengzhou@fudan.edu.cn

**Abstract**

Flash memory has become a ubiquitous solid-state memory device, it is widely used in portable digital devices, computers, and enterprise applications. The development of the information age has put forward higher requirements for memory speed and retention performance. Here, we demonstrate an ultrafast non-volatile memory based on $MoS_2$/h-BN/multi-layer graphene (MLG) van der Waals heterostructures, which has an atomic-level flat interface and achieves ultrafast writing/erasing speed (~20 ns), surpassing the reported state-of-the-art flash memory (~100 ns). The ultrafast flash memory could lay the foundation for the next-generation of high-speed non-volatile memory.

**Introduction**

Since the discovery of the floating-gate memory in 1967[1], flash memory has played a pivotal role in the field of non-volatile memory based on its operating principle. To overcome the physics and scaling limitations of the two-dimensional (2D) planar cell structure, 3D flash memory structures have been developed[2,3]. In the development of the era of big data, the emergence of cloud computing, artificial intelligence, virtual reality, and the Internet of Things has led to a higher demand for cell performance such as the programmed speed.

In recent years, the performance of the 2D materials transistors regulated by 1-nm gate electrodes has attracted much attention[4], which overcomes the short-channel effect and quantum confinement effect in the silicon-based transistors in the small footprint. Without hanging bonds on the material surface, van der Waals heterostructures show advantages in functional devices exploration[5-9]. The quest for new types of memories based on 2D materials has generated several marvelous ideas. Broadly speaking, these memories based on 2D materials could be divided into three main categories: volatile memory such as SRAM[10-12], quasi-non-volatile memory with a game between write speed and retention characteristics[13-15], non-volatile memory with the write speed of only milliseconds[16-24]. It is worth noting that the millisecond-level speed of non-volatile memory is far from meeting the needs of next-generation high-speed memory.

The high-quality interface of van der Waals heterostructures may push traditional memory to higher performance, such as improving the writing speed of flash memory. As one of the most extensively studied transition metal dichalcogenides semiconductors, $MoS_2$ possesses a high current ON/OFF ratio ($\sim 10^8$) in a single layer[25] and a bandgap range from 1.2 eV to 1.8 eV with the thickness decreases from bulk to single layer[26], which can function as channel materials in flash memory. h-BN can be used for insulator and tunneling layer due to its wide bandgap ($\sim 5.9$ eV)[27,28]. Multilayer graphene with high work function ($\sim 4.6$ eV), high density of states, and good thermal stability can be good choice for the floating-gate materials.[29,30].

Here, for the first time, we demonstrate an ultrafast flash memory based on $MoS_2$/h-BN/MLG van der Waals heterostructures. By utilizing this van der Waals heterostructures, a flash memory with 20 ns ultrafast writing/erasing speed and long retention time characteristics was achieved, which is faster than the reported state-of-the-art flash memory[31]. This ultrafast flash memory could pave the way for the next-generation high-performance memory applications.

**Results and Discussion**

Figure 1a, 1b is the schematic structure and the cross-sectional transmission electron microscope image of the $MoS_2$/h-BN/MLG heterostructures, respectively. Figure 1c shows the top view of the optical microscope photograph of our memory device on a low-resistance silicon substrate with a 285-nm-thick $SiO_2$. The stacking order of the heterostructures is marked from top to bottom by blue, white and red dashed lines representing channel $MoS_2$, tunneling layer h-BN, and floating-gate layer multilayer graphene, respectively. A floating-gate electrode was connected to the MLG layer to verify the memory characteristics. Cr/Au (10 nm/30 nm) as the source, drain, and floating-gate electrodes. The detailed device preparation methods and processes are provided in the Experimental Section and Supplementary Figure 1, respectively. The typical Raman spectra of the overlapping parts of the heterostructure is shown in Fig. 1e. Due to the relationship between the frequency shift and the thickness of the $MoS_2$

layered material[32], the difference between the two Raman modes, $E^1_{2g}$(386 cm$^{-1}$) and $A_{1g}$ (408 cm$^{-1}$) was 22 cm$^{-1}$, i.e., corresponds to the thickness of the MoS$_2$ was two-layer. The typical Raman shift of h-BN ($E_{2g}$ peak at 1366 cm$^{-1}$)[33] and multilayer graphene (G peak at 1580 cm$^{-1}$ and 2D peak at 2700 cm$^{-1}$)[34] were plotted in the Fig. 1d. The corresponding thickness of the h-BN and MLG were measured by atomic force microscope in Supporting Fig. 2a, 2b.

Figure 2a, 2d show the schematic diagrams and the electrical connections of the devices. The source-drain current was regulated by the back-gate and floating-gate, respectively. When the $|V_{bg, max}|$ applied to the back gate gradually increases, as shown in Fig.2b, the threshold voltage ($V_{th}$) has shifted, resulting in a distinct clockwise memory window at $V_{ds}$= 0.05 V. The memory window reaches 53 V when the $V_{bg}$ is swept back and forth between ±35 V. The extraction of the memory window for different $V_{bg,max}$ is shown in Supplementary Figure 3. As for the amount of charge stored in the floating layer MLG could be estimated from the expression[16]: n=($\Delta V \times C_{bl}$)/q, where $\Delta V$ is the difference of the threshold voltages shift, q is the electron charge (~1.6×10$^{-19}$ C), $C_{bl}$ is the capacitance between the back gate and floating-gate layer. The density of the stored electrons on the order of ~ 4×10$^{12}$ cm$^{-2}$, which in agreement with the values of the MLG as floating-gate layer that have been reported[29,30]. Since the electrons (holes) stored in the floating-gate layer which weaken the electric field strength of the $V_{bg}$, the channel current could only in two states as shown in the output characteristic curves of Fig.2c. The internal dashed box is an enlarged plot of the channel current $I_{ds}$ at 45 mV to 50mV. When the floating-gate electrode is grounded, the corresponding transfer characteristic curve (Supporting Fig. 4b) verifies that the accumulated electrons are stored in MLG. When the channel current is regulated by the floating-gate electrode (Fig. 2d), the h-BN acts as a dielectric layer. Figure 2e shows the transfer characteristic curves for the corresponding MoS$_2$ transistor. When the voltage is scanned from +6 V to -6 V and then to +6 V, no memory window appears. Due to a perfect interface of the heterojunction, the interface defect states are negligible.

Good ohmic contact was shown in the output characteristic curves of the MoS$_2$-FET in Fig. 2f.

The basic operation principle of the flash memory is to store electrons in a floating-gate layer, and the stored electrons change the $V_{th}$ of the device. Here, the ultrafast writing/erasing operation speed (~20 ns) is demonstrated based on our device in Figure 3. Firstly, a $V_{bg, pulse}$ = -30 V for 1 s pulse was applied to the back-gate to thoroughly erase the device to state 1, the corresponding $I_{ds}$-$V_{bg}$ was measured at $V_{ds}$= 0.05 V, as shown by the dotted line in Fig. 3a. Then, a $V_{bg, pulse}$ = +25 V for 20 ns pulse was applied to the bake-gate for the writing operation, and the corresponding curve of the $I_{ds}$-$V_{bg}$ was shown by the red solid line in the Fig. 3a. A clear rightward shift of the threshold voltage ($\Delta V_{th}$ >0 V) occurs due to the electrons of the MoS$_2$ tunneling into the MLG layer. As the pulse amplitude of the write operation is gradually increased to +35 V, the threshold voltage change is gradually saturated. When a $V_{bg, pulse}$ = +30 V for 1 s write operation applied to the Si, the corresponding $I_{ds}$-$V_{bg}$ curve was shown from the dashed line in Fig. 3b. Obviously, a different left shift of the threshold voltage ($\Delta V_{th}$ <0 V) appears after $V_{bg, pulse}$ = -20 V, -25 V, -30 V for 20 ns erase operation. The successful write and erase of ultrafast pulses of 20 ns gives superior performance to the reported state-of-the-art non-volatile flash memory, including organic flash memory[35-37], 2D flash memory[16,18,22-24,30,38-42] and silicon-based flash memory[31,43,44], the corresponding statistical comparison is shown in Supporting Figure 13.

The variation of the $\Delta V_{th}$ (liner fitting of the corresponding $I_{ds}$-$V_{bg}$ in Supporting Figure 5) for flash memory devices with different pulse widths was summarized in Fig. 3c and the test conditions are fixed at $V_{ds}$= 0.05 V, in vacuum at room temperature. By quantifying the write/erase pulse width and pulse amplitude, the memory device has the potential for multi-level storage in the future.

The breakdown field strength of the tunneling layer (h-BN) was measured to be 7.43 MV/cm (Supplementary Fig. 6d), which in agreement with previous reports[45], thus direct tunneling mechanism[24] play an important role just at the maximum pulse voltage of ±35 V. However, the leakage current through the 2D dielectrics is different from the conventional 3D dielectrics, which is not well understood. The main reason is that some factors which do not exist in conventional 3D dielectrics may play a key role, such as the van der Waals gap between the 2D materials and the insulator, which may play a significant role in reducing the tunneling current[46]. The ultrafast speed memory in this work may benefit from the flat and clean two-dimensional interface increased the tunneling probability, suitable energy band structure, or accelerated write speed assisted by electron injection at the control gate and source terminal, more exploration needs to be done in the future.

As for the ultrafast memory characteristics, it is shown in Fig. 3d. When the device stays in any initial state, at 60 s, a $V_{bg,\ pulse}$ = -30 V for 50 ns erase operation put the channel current to state 1 (~ 5 μA), then stable in a high state until 160 s, at which point a $V_{bg,\ pulse}$ = +30 V of 50 ns write operation causes the device to plunge to the state 0 (~ 1pA). The corresponding pulse widths (Full Width at Half Maximum, FWHM) are 48 ns, 46 ns, respectively (in Supporting Fig. 8c, 8g). To exclude the influence of the interfacial state of the 2D material with the blocking layer, we fabricated a single $MoS_2$ back-gate field-effect transistor device, however, the fastest write speed of the memory based on $MoS_2$-FET is only in the order of milliseconds (in Supplementary Fig. 8d). To investigate the stability and the tunability of the flash memory device, the dynamic behavior was demonstrated. By applying -30 V, 0 V, and +30 V of $V_{bg}$, 20 ms duration, the dynamic behavior of the device was achieved at $V_{ds}$= 0.05 V. Figure 3e shows the switching state after 1185 erase and write cycles, and the channel current memory state-1/state-0 still $10^4$ under the read condition of $V_{ds}$= 0.05 V, $V_{bg}$= 0 V, exhibiting stable fatigue characteristics. The blue dashed box on the right is a real-time memory state in one of the cycles zoomed in Fig. 3e.

For the memory performance at different temperatures, we demonstrated the transfer characteristic of the device after $V_{bg,\ pulse}$= 30 V for 50 ns write/erase operation (dashed lines in Fig. 4a) and $V_{bg,\ pulse}$= -30 V for 50 ns erase operation (dashed lines in Fig. 4b) at 20 K to 300 K, respectively. Before performing the ultrafast write operation, a pulse of $V_{bg}$= -25 V/+25 V for 1 s was applied to the back gate, erased/write the memory to state 1/state 0. The corresponding transfer characteristic of the memory was shown by the solid lines in Fig. 4a/Fig. 4b. In Fig. 4a, 4b, the threshold voltage shifts obviously to the left with increasing temperature, due to the change of the charge concentration caused by the associated metal-insulator transition in the $MoS_2$[47]. Especially, this transition manifests itself as a crossing-over between $I_{ds}$ versus $V_{bg}$ curves obtained at different temperatures (in Supporting Figure 9) indicating two different regimes. Nonetheless, the $\Delta V_{th}$ is highly stable after the write and erase operation at different temperatures, by linearly fitting the transfer characteristic curves in Fig. 4**a** and 4**b**. As an indicator of the non-volatile memory, data retention characteristics are essential. By applying a positive/negative pulse ($V_{bg,\ pulse}$ = +30 V/-30 V for 50 ns) to the back-gate Si, we measured the $\Delta V_{th}$ at different time intervals (linear fitting the $I_{ds}$-$V_{bg}$ curves in Supporting Figure 10), as shown in the Fig. 3d. The relative threshold difference changes to 50.6% and 50.4% of its initial difference at $3.15\times10^8$ s, respectively, read at $V_{ds}$=0.05 V, at the room temperature. A good data retention performance was achieved in our device.

Subsequently, we investigated the influence of the thickness of the tunnel layer on the ultra-high-speed performance. Devices with 7.5 nm-thick-BN and 18 nm-thick BN of the memory devices were fabricated, respectively. The thickness of the corresponding tunneling layer is shown in Supporting Fig. 2d, 2f. Figure 5a shows the $I_{ds}$-$V_{bg}$ curves of the 7.5 nm-thick-BN memory with forward and reverse control gate voltage (between +19 and -19 V) scanning on the back gate at $V_{ds}$=0.05 V, which exhibits significant hysteresis (25 V) due to the aforementioned charge injection and trapping in the MLG floating-gate. Based on the test method illustrated in Fig. 3a, 3b, the histogram of the threshold voltage difference for different pulse widths at different voltages is shown in Figure 5c. The statistical results extracted from the $I_{ds}$-$V_{bg}$ curves (in Supporting Figure 11) shows that the process of charge tunneling and charge accumulation within 100 ns could by selecting the appropriate voltage amplitude and voltage time interval to obtain the corresponding threshold voltage. The retention performance of the 7.5 nm-thick h-BN memory is shown in Fig. 5c. By applying a $V_{bg,\ pulse}$ =+ 21 V for 50 ns erase pulse to the Si back gate, the device could be pushed to a high current state (~3.3 μA) and maintained to 0.7 μA till 1000 s, whereas a $V_{bg,\ pulse}$ =-19 V for 50 ns write pulse causes it to the low current state abruptly (~1 pA) and could not maintain the lowest current value, only be maintained to 30 nA till 1000 s, read at $V_{bg}$= 0 V, $V_{ds}$=0.05 V. The result shows that the 7.5 nm-thick h-BN memory could not maintain a better channel current state 1/0 ratio because of the increased leakage current than a reasonable thicker h-BN insulator (~10 nm).

When the thickness of the h-BN was increased to 18 nm, the transfer characteristics of the device with forward and reverse back gate voltage sweeping are also carried out (Fig. 5d), which exhibits a smaller hysteresis due to the 18 nm-thick h-BN weakens the tunneling probability. By linearly fitting the corresponding transfer characteristics of the 18 nm-thick h-BN memory device, the histogram of its corresponding threshold voltage difference variation is shown in Fig. 5e (extracted from the $I_{ds}$-$V_{bg}$ curves in Supporting Figure 12). By applying a $V_{bg, pulse}$ = -40 V for 1 s pulse to the back gate to put it in a fully erased state, then a +40 V write pulses of different pulse widths were applied to the Si, the fastest write speed is just 10 μs (the green histogram area in the Fig. 5e). The corresponding retention performance was shown in Fig. 5f. As shown in Fig. 5f, after application of voltage pulse (±40 V, pulse width 1 s) to the back-gate, the device exhibits a stable memory state 1/0 ratio (~10), read at $V_{bg}$= 0 V, $V_{ds}$=0.05 V. When the read condition is at $V_{bg}$= -10 V, $V_{ds}$=0.05 V, the channel current of the state 1 gradually decreases, due to the leakage current caused from the direct tunneling of the h-BN under the DC bias voltage. Therefore, the choice of an appropriate tunneling layer thickness is a prerequisite for achieving ultrafast non-volatile memory, if, only the ultrafast speed properties of the structure are needed for the exploration of other functional devices, which will be another different story.

**Conclusion**

By using van der Waals heterostructures we have demonstrated an ultrafast non-volatile memory. Based on a suitable thickness of the tunneling layer (h-BN), this type of memory has excellent properties of writing/erasing time as low as ~20 ns, channel current ration of state 1/state 0 ~$10^6$, and non-volatile retention performance. It should be noted that the upper limit of the writing/erasing speed is limited to the instrument response, and faster memory speed would be further explored. We anticipate that the van der Waals ultrafast flash memory will support the next-generation of high-speed non-volatile memory applications.

**Experimental Section**

Firstly, highly doped P-type silicon with 285 nm $SiO_2$ was selected as a substrate. $MoS_2$, h-BN, and graphene nanosheets were obtained from the commercially available crystals (2D Semiconductors, Inc.) by Scotch-tape micromechanical cleavage technique, respectively[48]. Secondly, $MoS_2$ and h-BN were transferred to a suitable position on the top of graphene at one time by the wet transfer method with the help of a water-soluble layer (PVA)[27]. The tunneling layer, h-BN, should be transferred to a suitable location to avoid shortening between the channel ($MoS_2$) and the MLG. Thirdly, the electrodes were patterned by electron beam lithography (EBL) using polymethyl methacrylate (PMMA) (AR-679.04) polymer. Finally, Cr/Au (10 nm/30 nm) electrodes were fabricated by thermal evaporation and lift-off steps. The corresponding experimental flow chart was shown in supporting S1.

The electrical performance of the devices was measured by Agilent B1500 semiconductor parameter analyzer. The electronic test platform was the Cascade summit 11000 type in the room temperature atmospheric condition. The electrical test platform for low-temperature conditions was on a Lake Shore probe station. The thickness of the nanosheets was measured by MFP-3D Origin$^+$.

**Acknowledgments**

This work was supported by the National Natural Science Foundation of China (61925402, 61851402 and 61734003), Science and Technology Commission of Shanghai Municipality (19JC1416600, National Key Research and Development Program (2017YFB0405600), Shanghai Education Development Foundation and Shanghai Municipal Education Commission Shuguang Program(18SG01).

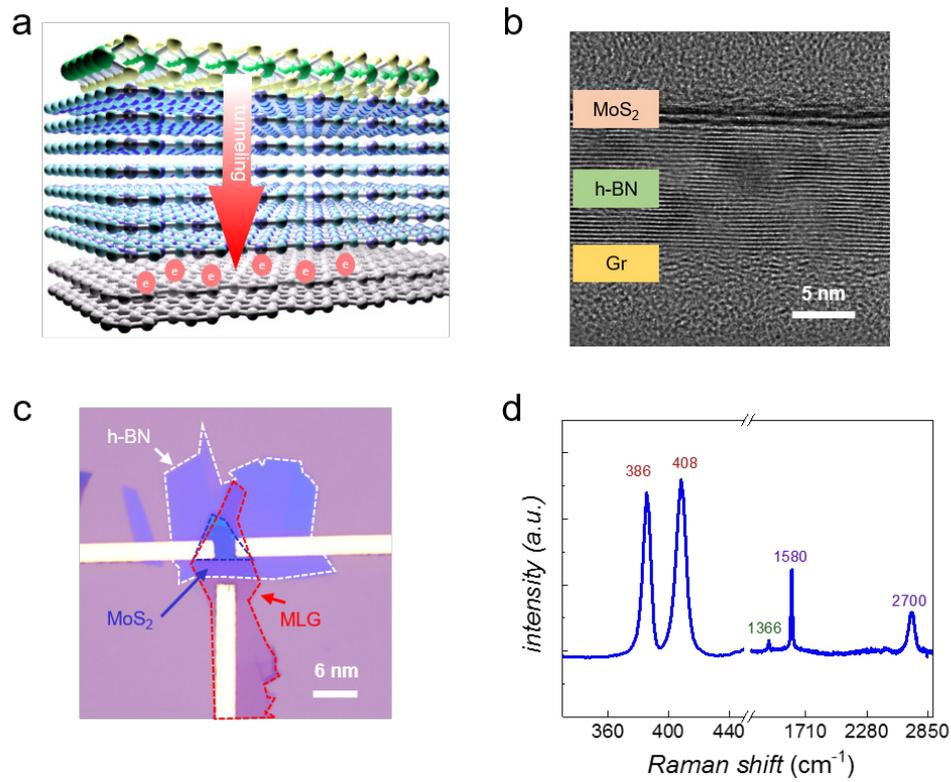

**Figure 1. Van der Waals heterostructures flash memory with atomic-level flat interface. a**, Schematic structure of the MoS$_2$/h-BN/MLG heterostructures. The MoS$_2$, h-BN, and graphene act as the channel material, tunneling layer, and floating-gate layer, respectively. **b**, The cross-sectional transmission electron microscope image of the device. The interfaces of different layers are flat and clean. The thickness of the h-BN is about 10.5 nm. The scale bar is 5 nm. **c**, Optical micrographs of the device. The overlapping functional layers are surrounded from top to bottom marked by the blue, white and red dashed lines for MoS$_2$, h-BN, and graphene, respectively. The width and length of the MoS$_2$ channel are 3.3 μm. Scale bar is 6 μm. **d**, The typical Raman spectra of the MoS$_2$, h-BN, and graphene in the overlapping region.

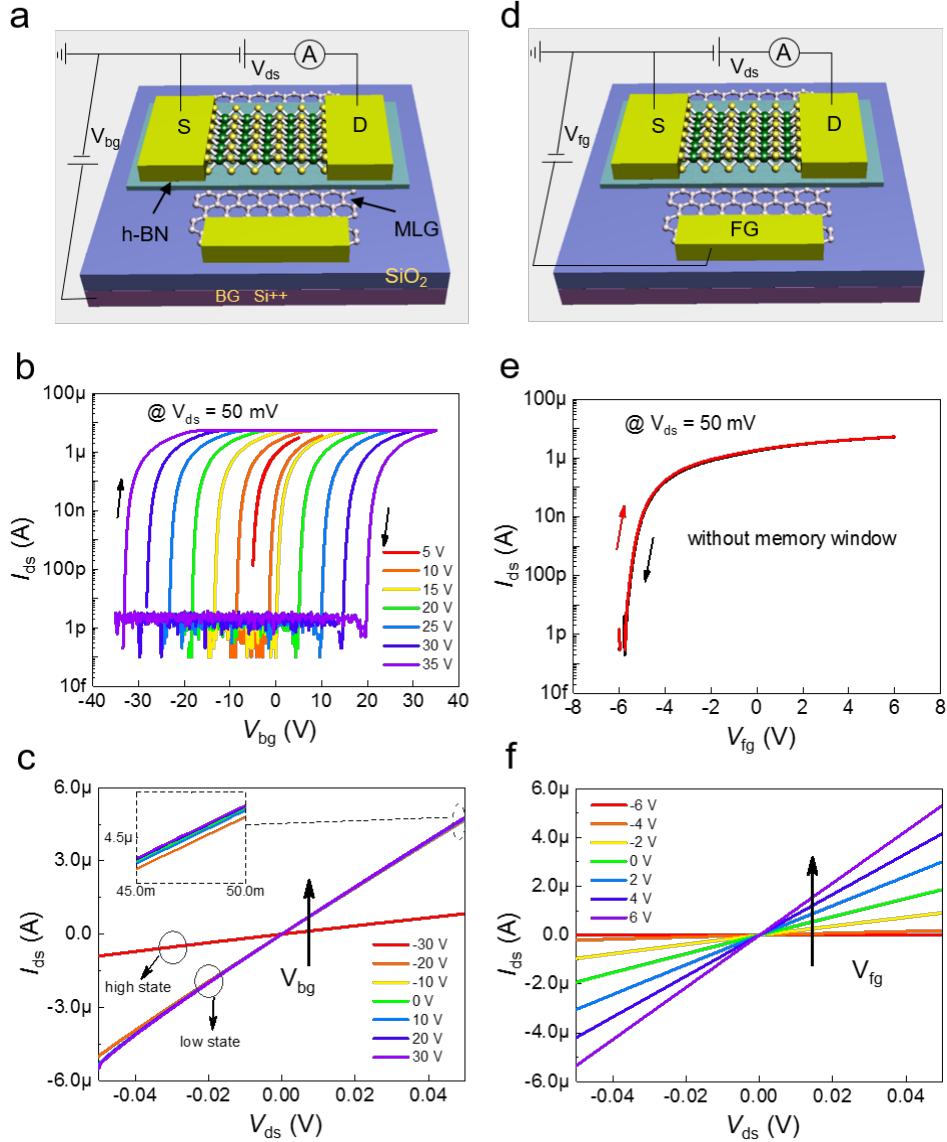

**Figure 2. Transfer and output characteristic curves of the flash memory device. a**, The device structure diagram modulated by the back-gate electrode (low-resistance, highly doped silicon) with 285 nm SiO$_2$ blocking layer. **b**, Transfer characteristics of the device controlled by back-gate. The shift of the threshold voltage leads to the generation of the memory window, and the memory window value can reach to 53 V at $V_{ds}$= 0.05 V and $V_{bg}$ = 35 V. **c**, Output characteristic curves of the device controlled by back-gate, due to the accumulation of tunneling charges in the graphene, so there are only two clear and distinct states, i.e., the high and low states marked by black circles. **d**, The device structure diagram modulated by the floating-gate electrode. The floating-gate electrode is in direct contact with the graphene and stays away from the MoS$_2$ and h-BN. **e**, Transfer characteristics of the MoS$_2$-FET under the control of graphene floating-gate, graphene acts as electrode and h-BN acts as gate dielectric layer. The excellent interlayer interface of the van der Waals heterostructures allow the device to scan back and forth without memory windows at ±6 V, showing negligible interface charge states. **f**, Output characteristic curves of the MoS$_2$ field-effect transistor, showing a good ohmic contact.

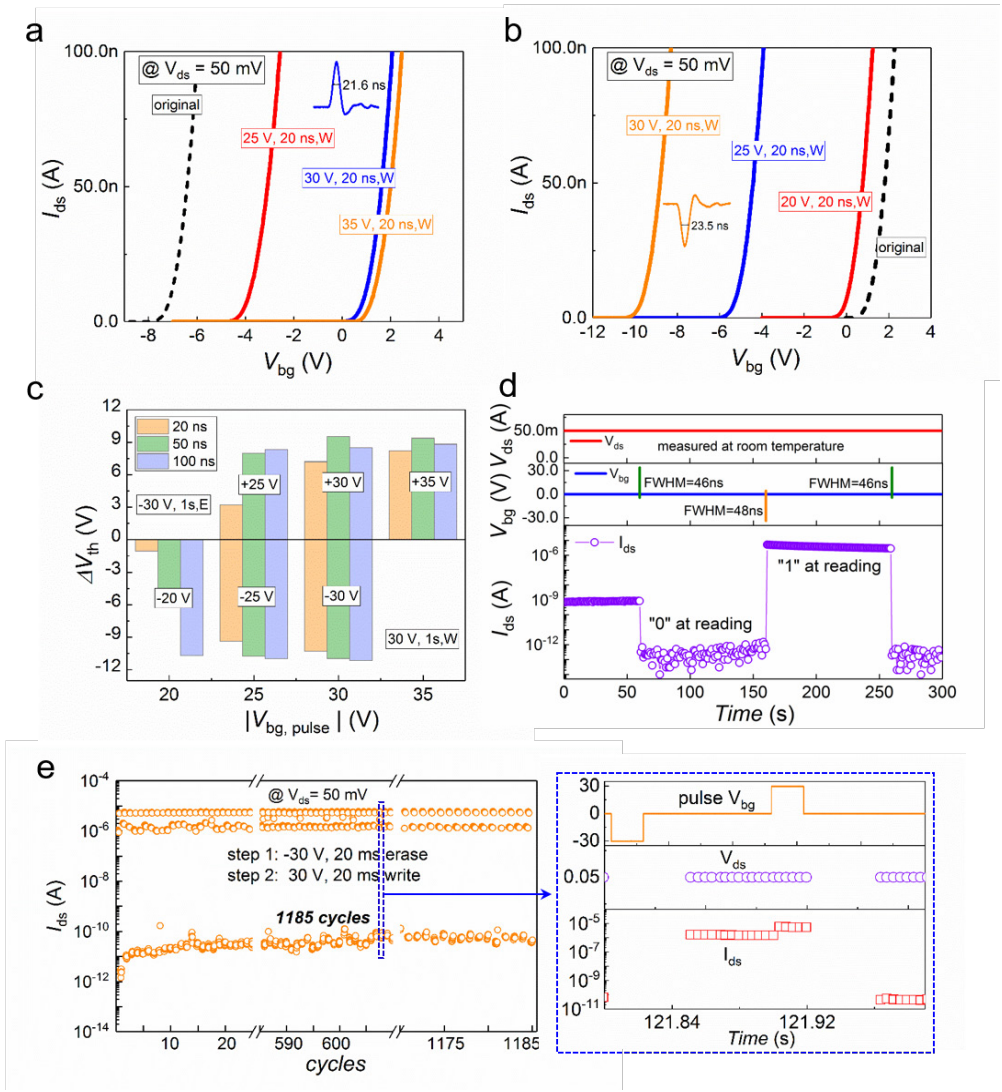

**Figure 3. Performance characterization of the ultrafast flash memory device. a**, **b** The transfer characteristics of the NVM device. By applying different positive pulses (amplitudes 25 V, 30 V, and 35 V, pulse duration 20 ns) programmed the device, the device shows a significant right shift in the threshold voltage. When the device was erased by negative pulses (amplitudes -20 V, -25 V, and -30 V, pulse duration 20 ns), a significant left shift in the threshold voltage can be observed. The actual write/erase pulse was generated by B1500 (FWHM was 21.6 ns/23.5 ns). **c**, A summary of the threshold voltage changes at room temperature for different pulse amplitudes and pulse widths. **d**, The variation of channel current values after the write (+30 V, 46 ns) and erase (-30 V, 48 ns) operations. A large memory state-1/state-0 ratio was observed which exceeded $10^6$, reading at $V_{ds}$= 0.05 V and $V_{bg}$=0 V. **e**, Endurance of the device for 1185 cycles of the writing/erasing pulse (+30 V, 20 ms / -30 V, 20 ms). The dashed blue box on the right is the real-time change of the channel current value for one cycle.

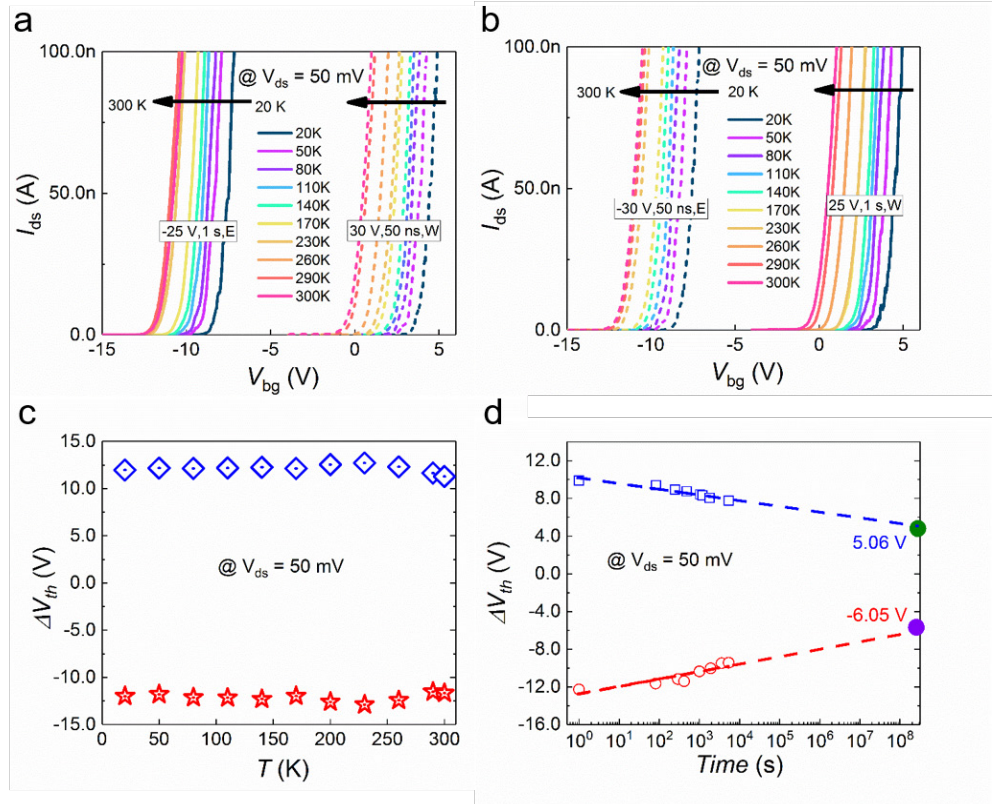

**Figure 4. Stable storage characteristics at variable temperatures. a**, The ultrafast writing characteristics under different temperature. The solid lines represent the transfer characteristic curves at different temperatures after applying a $V_{bg}$= -25 V for 1 s erasing pulse, and the dashed lines represent the transfer curves of the device after applying a $V_{bg}$= +30 V for 50 ns pulse. **b**, The ultrafast erasing characteristics under different temperature. The solid line represents the transfer characteristic curves at different temperatures after applying a $V_{bg}$= +25 V for 1 s duration writing pulse, and the dashed line represents the transfer curves of the device after applying a $V_{bg}$= -30 V for 50 ns duration erasing pulse. **c**, The threshold voltage shift $\Delta V_{th}$ at different temperatures. The data was extracted by linearly fitting to the transfer characteristic curves in Fig. 4**a**, **b**. **d**, The retention time of the threshold voltage difference at room temperature. The writing and erasing pulses are set to +30 V for 50 ns and -30 V for 50 ns, respectively.

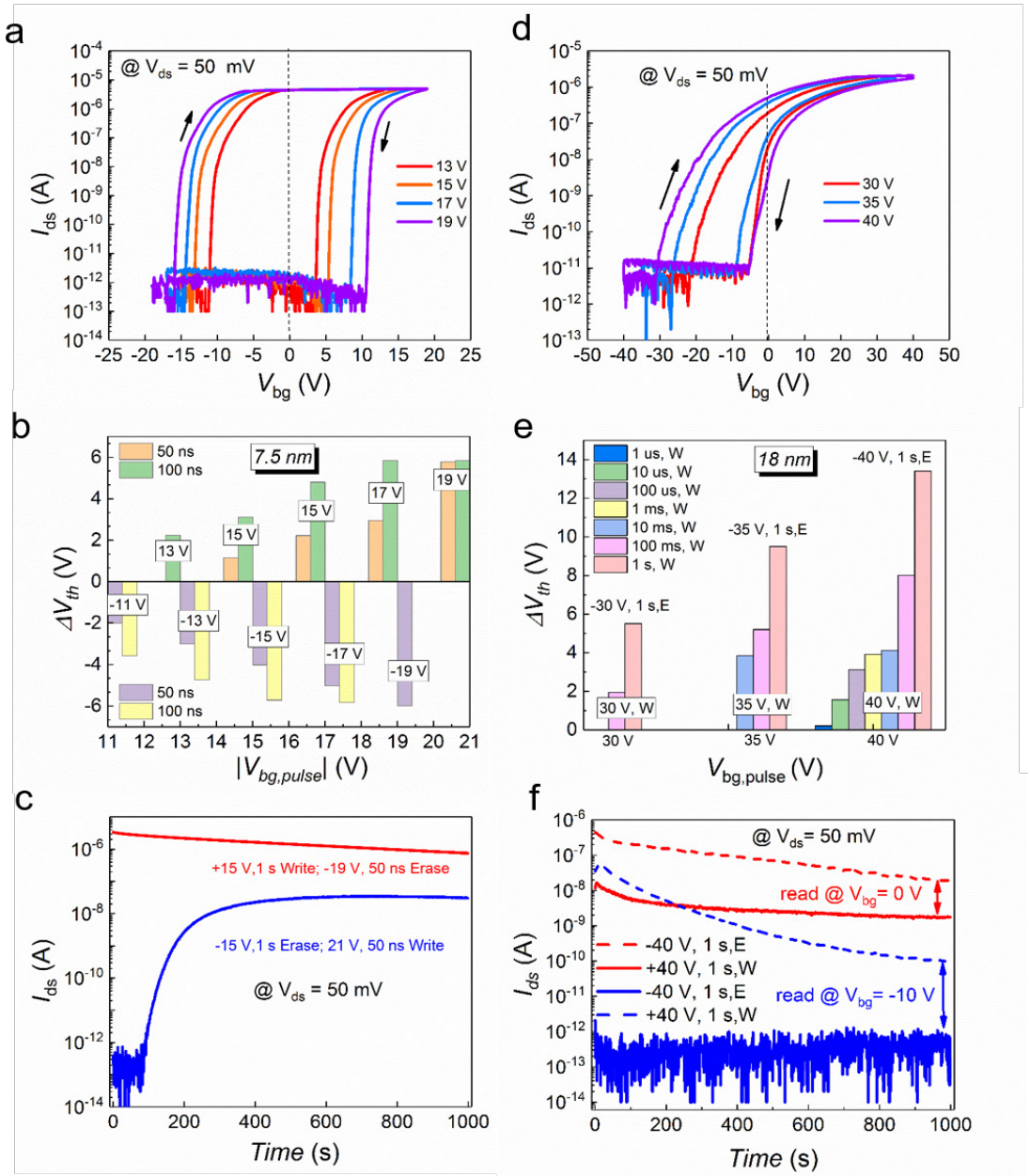

**Figure 5. Memory performance of devices with different tunneling layer thickness. a**, Transfer characteristics curves of the device with 7.5 nm h-BN tunneling layer. **b**, The difference value of the threshold voltage in the 7.5 nm h-BN memory device. **c**, Retention performance of the memory device with 7.5 nm h-BN. The writing/erasing pulse is +21 V, 50 ns /-19 V, 50 ns and the readout condition is $V_{bg}$= 0 V, $V_{ds}$ = 0.05 V. **d**, Transfer characteristics of the device with 18 nm h-BN tunneling layer. **e**, The threshold voltage shift of the 18 nm h-BN memory device. The fastest write pulse is 10 μs at $V_{bg}$= +40 V. **f**, Retention performance of the memory device with 18 nm h-BN after applying $V_{bg}$=±40 V for 1 s. The blue and red curves represent read conditions for $V_{bg}$=-10 V, $V_{ds}$=0.05 V and $V_{bg}$=0 V, $V_{ds}$= 0.05 V, respectively.

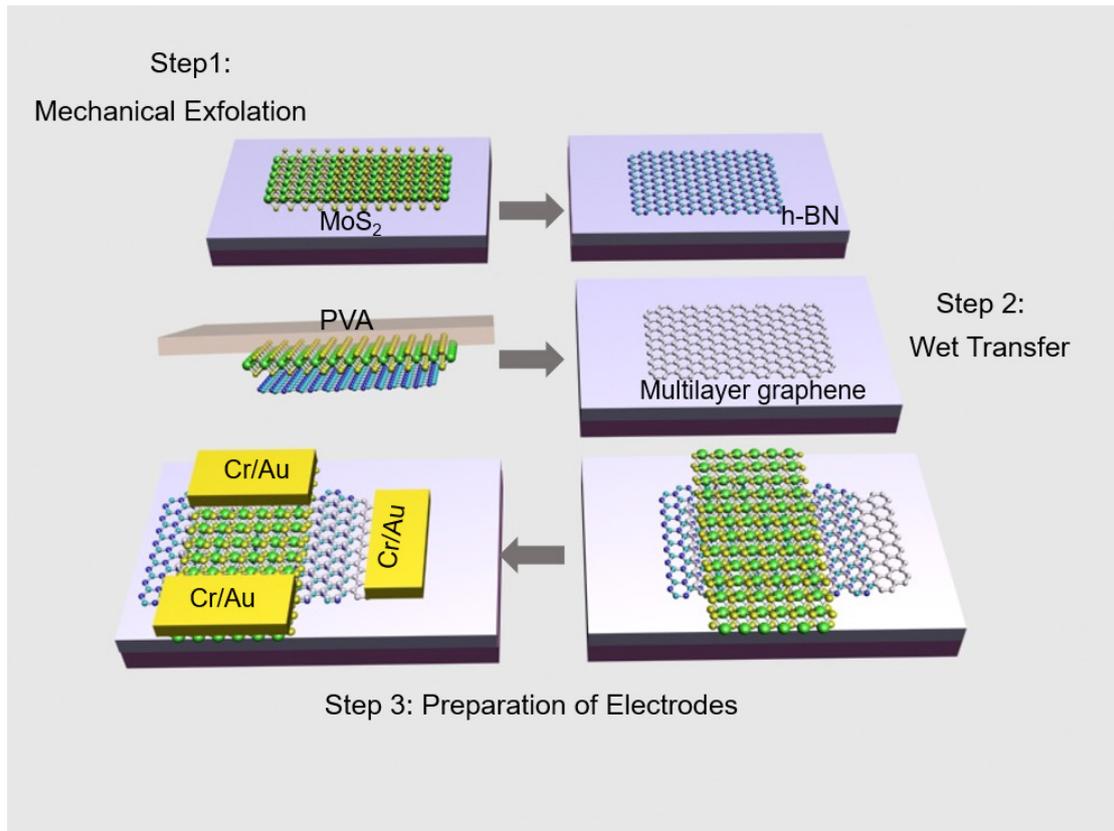

**Supplementary Figure 1. Schematic diagram of the preparation flow of the device.**

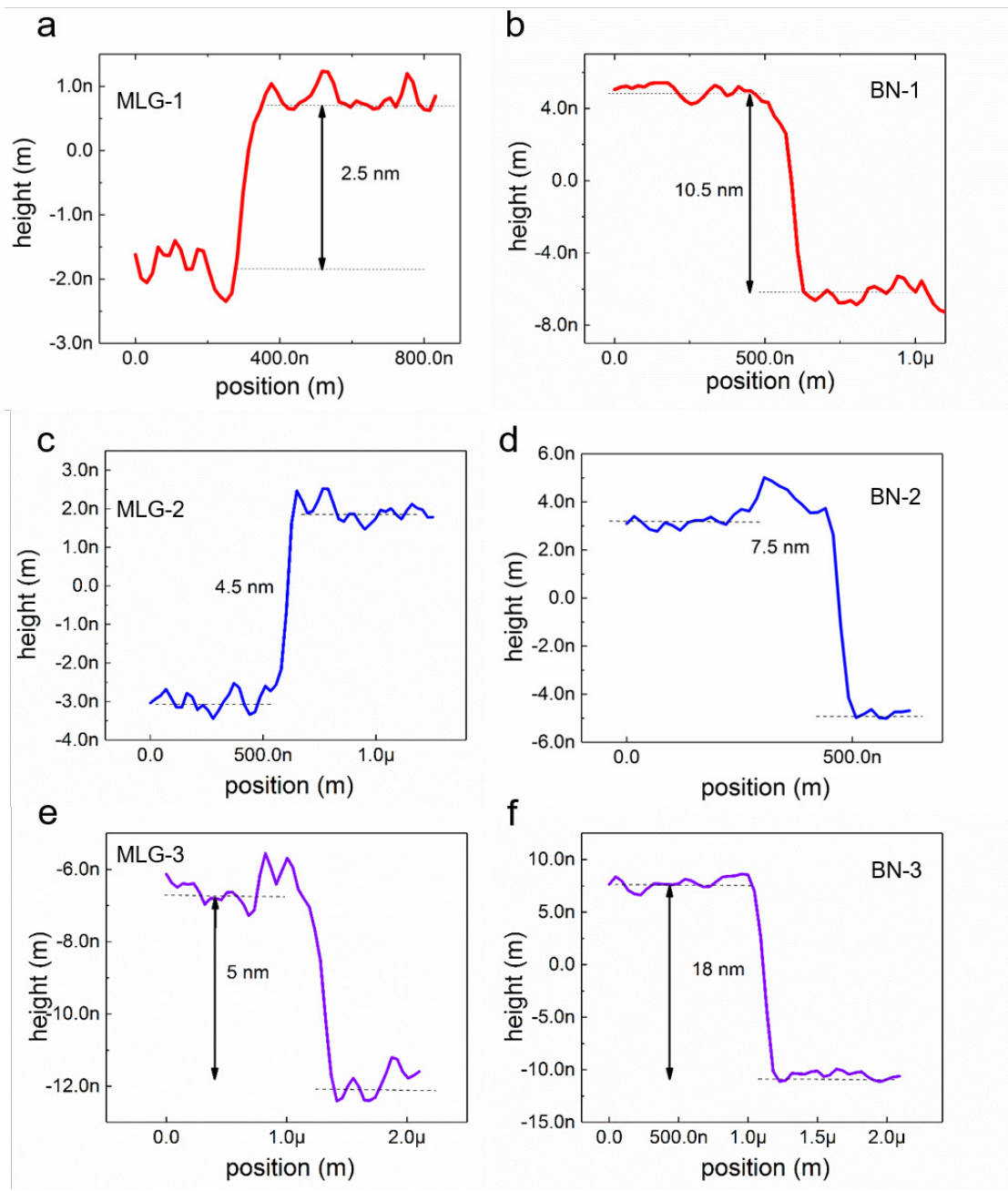

**Supplementary Figure 2. a, c, e,** The thickness of the floating layer in the different memory devices. **b, d, f,** The thickness of the tunneling layer in the different memory devices.

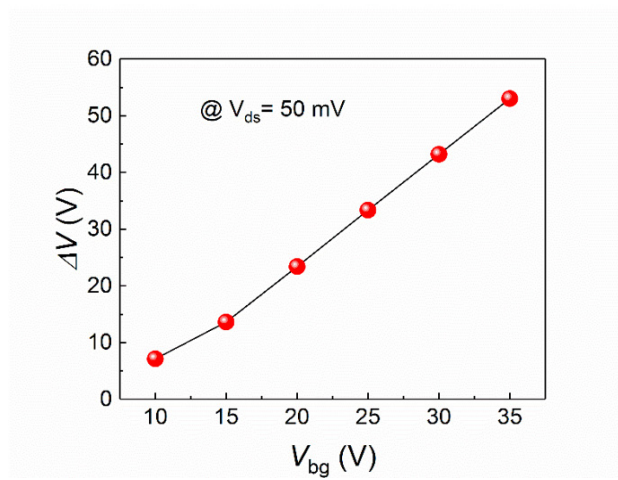

**Supplementary Figure 3. Extraction of the memory window *vs* maximum value of $V_{bg}$.** The memory window increases from 7.15 V to 53 V.

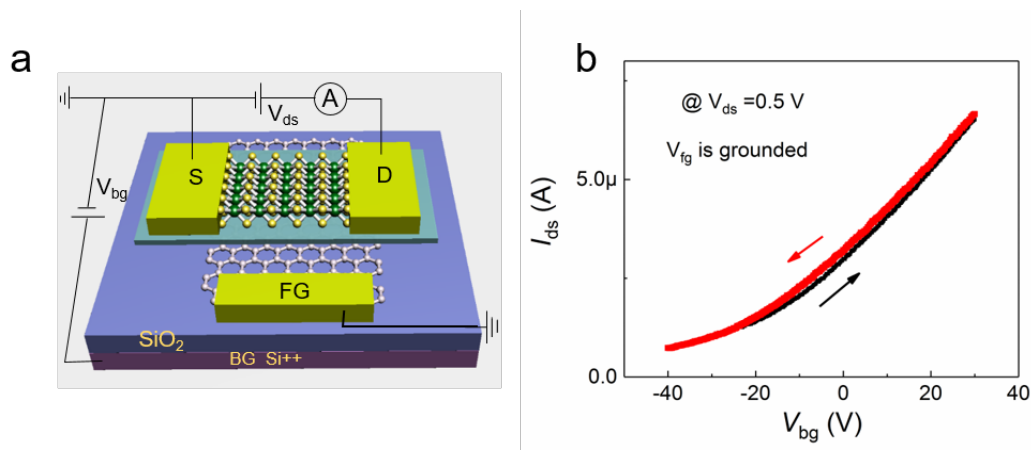

**Supplementary Figure 4. a,** The structure of the device which modulated by back-gate electrode, while the floating-gate electrode was grounded. **b**, The corresponding transfer characteristic curve of the memory device showing no memory window, reading at $V_{ds}$= 0.5 V.

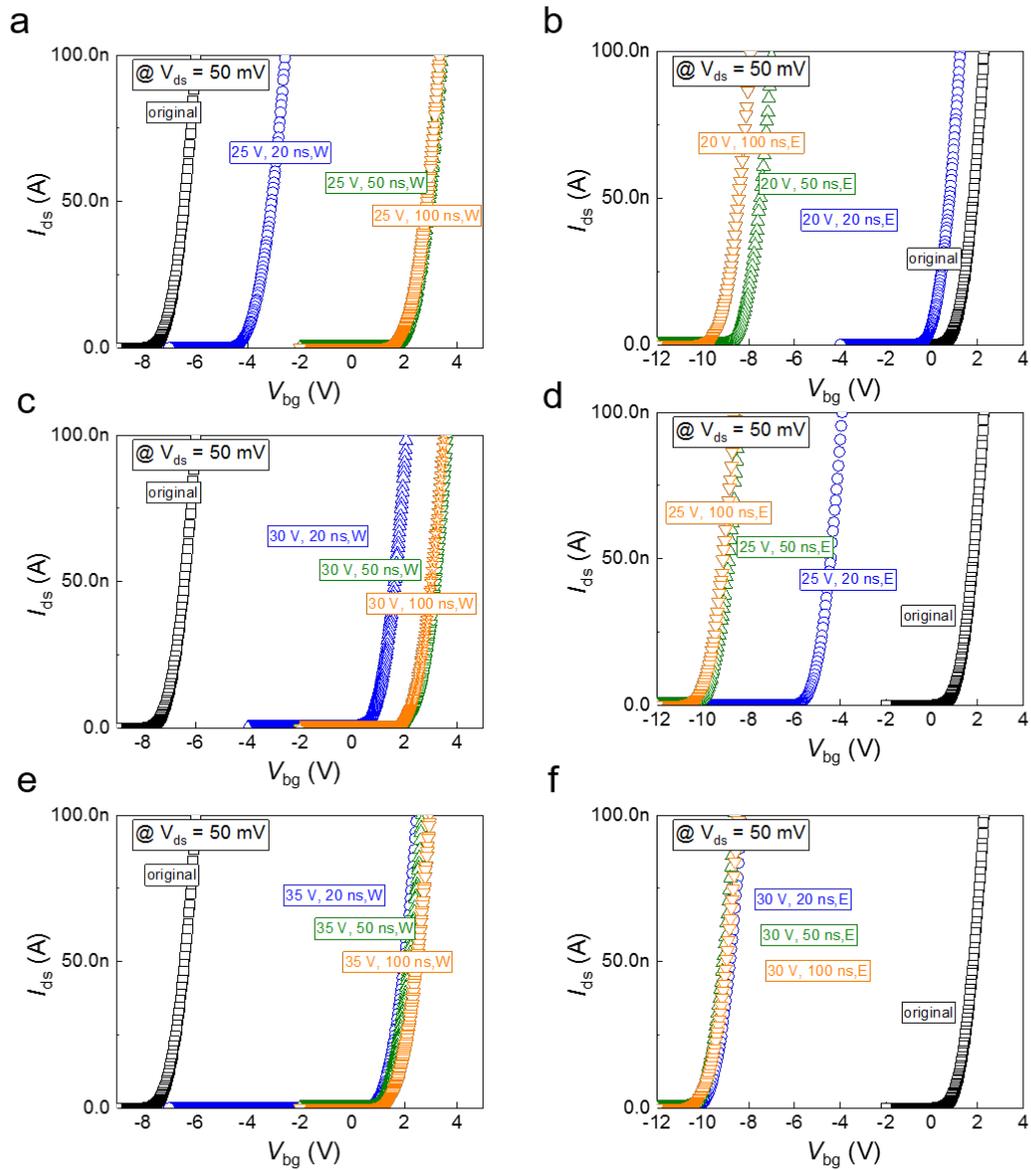

**Supplementary Figure 5. Transfer characteristic curves for different writing/erasing pulse amplitudes and pulse widths. a, c, e,** A $V_{bg,pulse}$= -25 V for 1 s duration erase operation put the device in the original state, the blue, orange and green curves representing the $I_{ds}$-$V_{bg}$ curves after applying a $V_{bg, pulse}$= +25 V (+30 V, +35 V) for 20, 50, 100 ns write pulse, respectively. **b, d, f,** A $V_{bg,pulse}$= +25 V for 1 s duration write operation put the device in the original state, the blue, orange and green curves representing the $I_{ds}$-$V_{bg}$ curves after applying a $V_{bg,pulse}$= -20 V (-25 V, -30 V) for 20, 50, 100 ns write pulse, respectively.

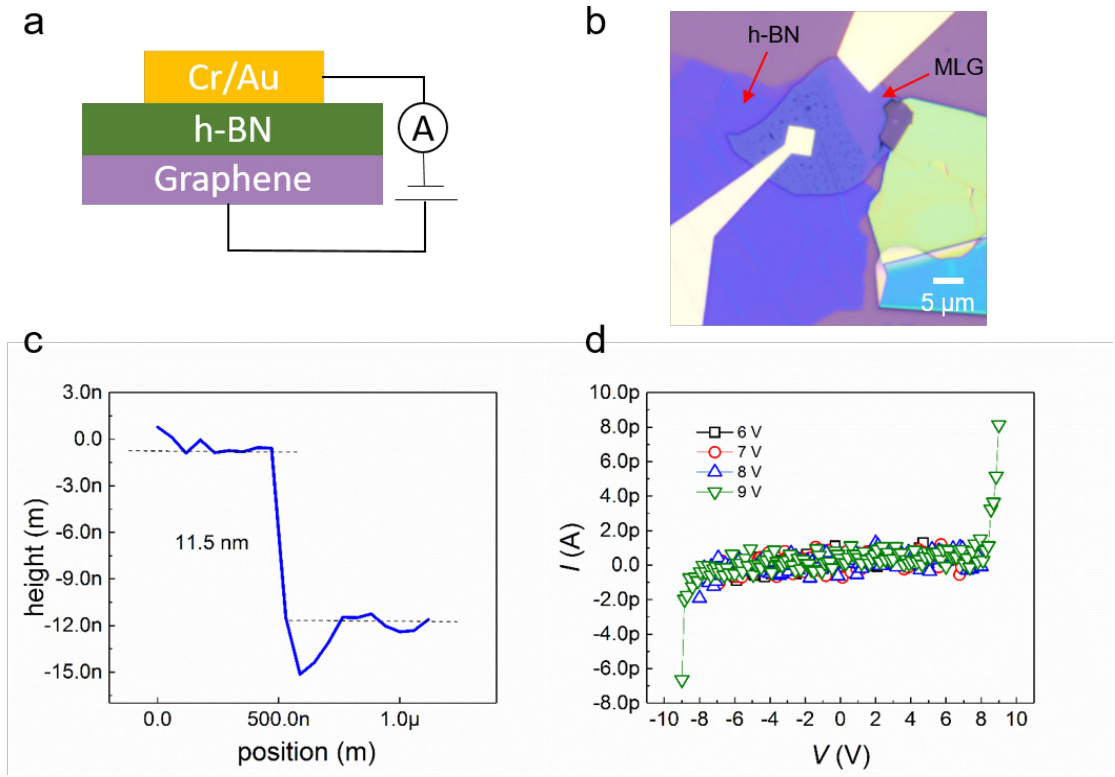

**Supplementary Figure 6. Tunneling current measurement in the h-BN. a**, Structure diagram of the Cr/h-BN/MLG device. **b**, Corresponding optical microscope photograph of the device. Scale bar is 5 μm. **c**, The thickness of the h-BN is determined by atomic force microscopy to be 11.5 nm. **d**, The tunneling current curves between h-BN at different voltages, from which Fowler-Nordheim tunneling occurs at 8.5 V.

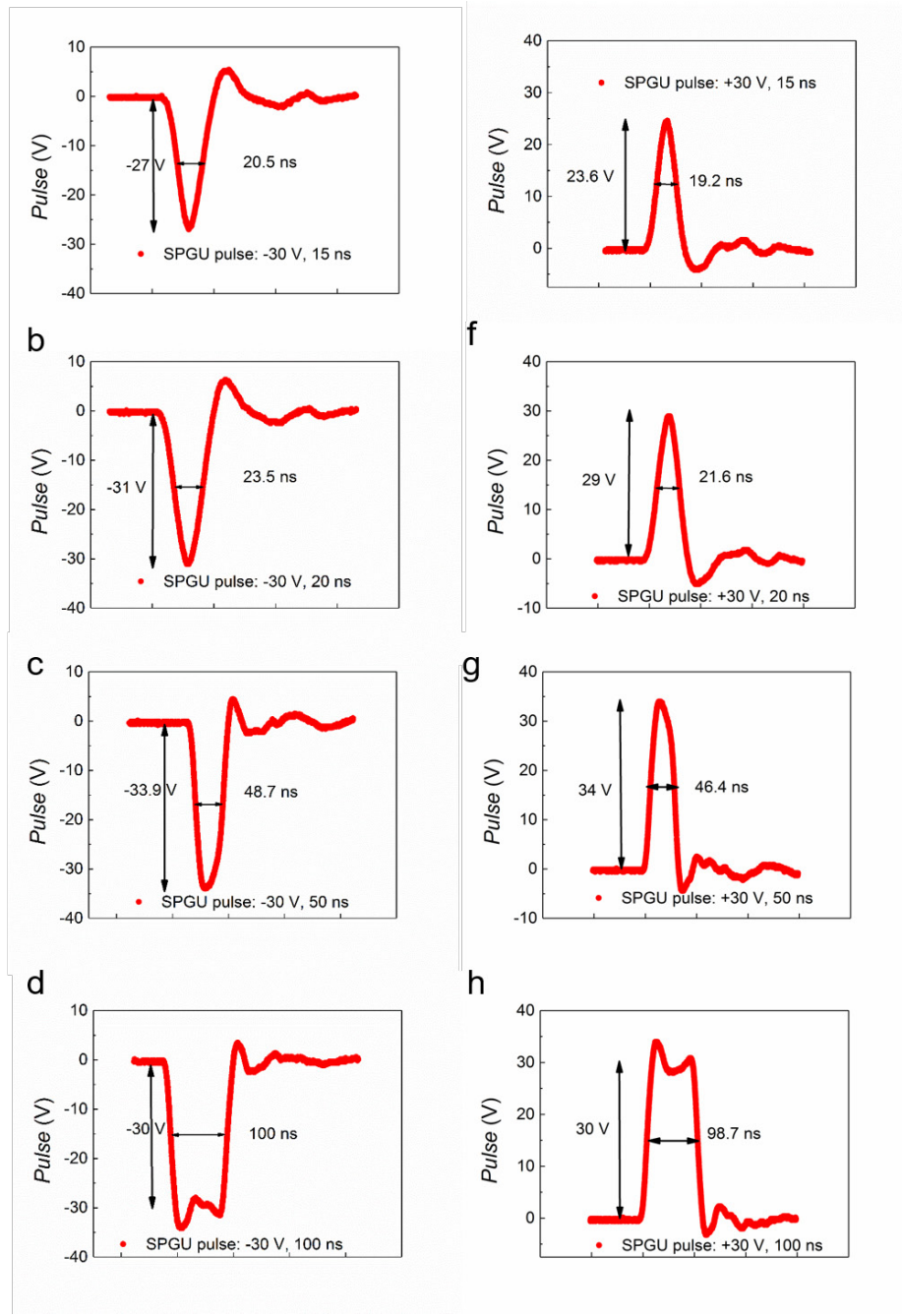

**Supplementary Figure 7. a**, **b**, **c**, **d**, Pulses with -30 V voltage amplitude and pulse width of 15 ns, 20 ns, 50 ns, and 100 ns generated by B1500 respectively and captured by oscilloscope MSO54. The corresponding effective duration of the pulse (FWHM, full width at half maximum) are 20.5 ns, 23.5 ns, 48.7 ns and 100 ns, respectively. **d**, **e**, **f**, **g**, Pulses with +30 V voltage amplitude and pulse width of 15 ns, 20 ns, 50 ns, and 100 ns generated by B1500, respectively. The corresponding FWHM are 19.2 ns, 21.6 ns, 46.4 ns and 98.7 ns, respectively.

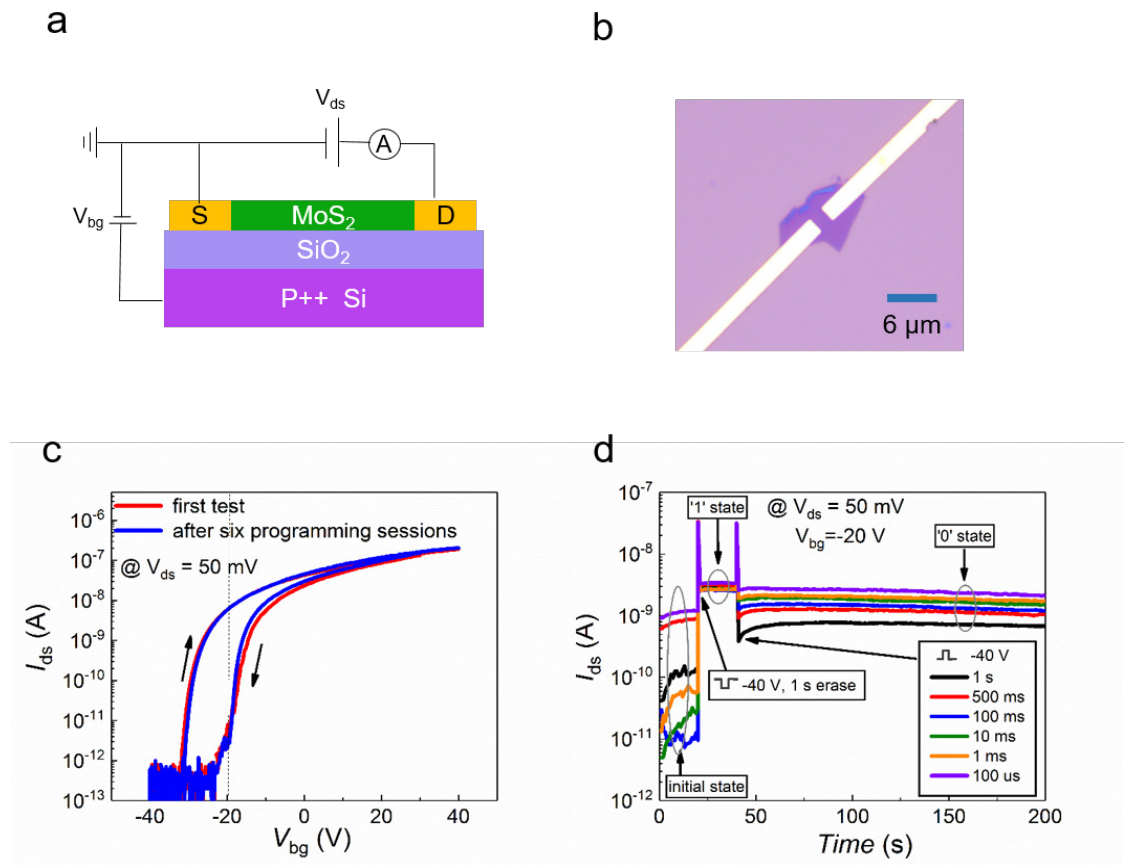

**Supplementary Figure 8. Memory performance of MoS$_2$-FET regulated by back-gate electrode**. **a**, Structure diagram of the MoS$_2$-FET device. **b**, Optical microscope photograph of the MoS$_2$-FET device. Scale bar is 6 μm. **c**, Transfer characteristic curves of the MoS$_2$-FET device before and after writing and erasing operations, reading at V$_{ds}$=0.05 V. A stable clockwise memory window mainly due to the presence of interfacial charge capture state as well as the channel defect state. **d**, A $V_{bg}$= -40 V for 1 s erase operation was applied to the Si to erase the device into a high current state (state 1), followed by +40V, different pulse width write pulse to operate it into a low current state (state 0), which showed that the fastest write speed was in the order of milliseconds in our experimental test (readout condition, $V_{bg}$= -20 V, $V_{ds}$= 0.05 V).

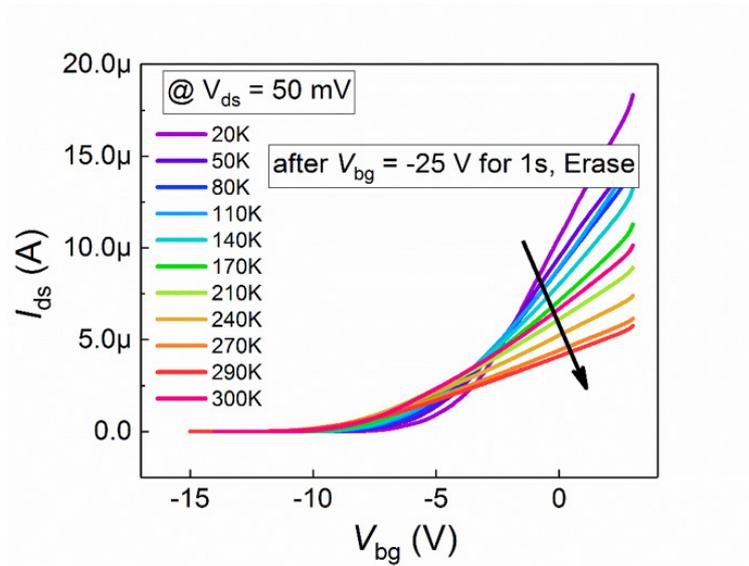

**Supplementary Figure 9**. Transfer curves corresponding to the erase state (applied by a $V_{bg,pulse}$=-25 V for 1s duration erased operation) of the memory at different temperatures.

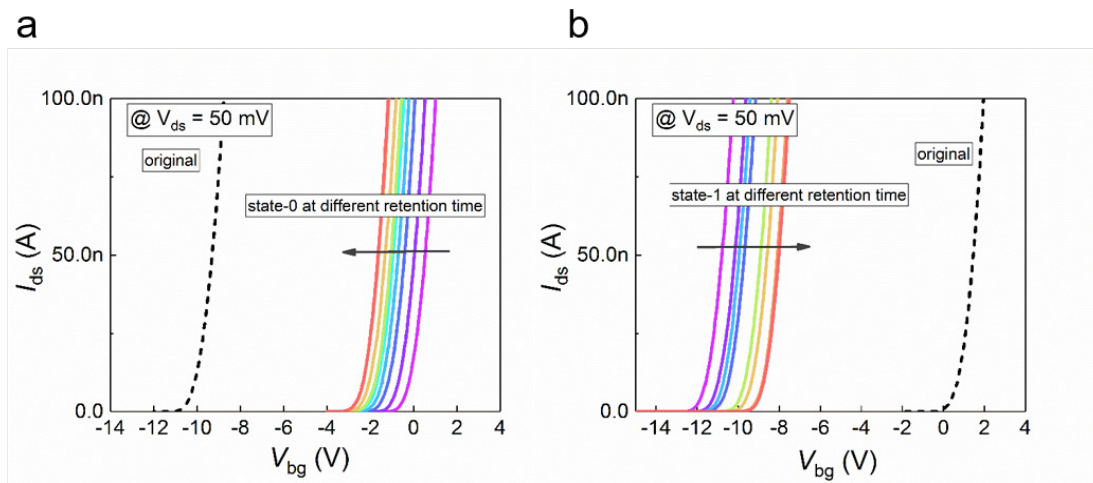

**Supplementary Figure 10**. **a**, Transfer characteristic curves of the 10.5 nm-thick h-BN flash memorydevice (after a $V_{bg,pulse}$=+30 V for 50 ns duration writed operation) at different time intervel. **b**, Transfer characteristic curves of the flash memorydevice (after a $V_{bg,pulse}$=-30 V for 50 ns duration erased operation) at different time interval .

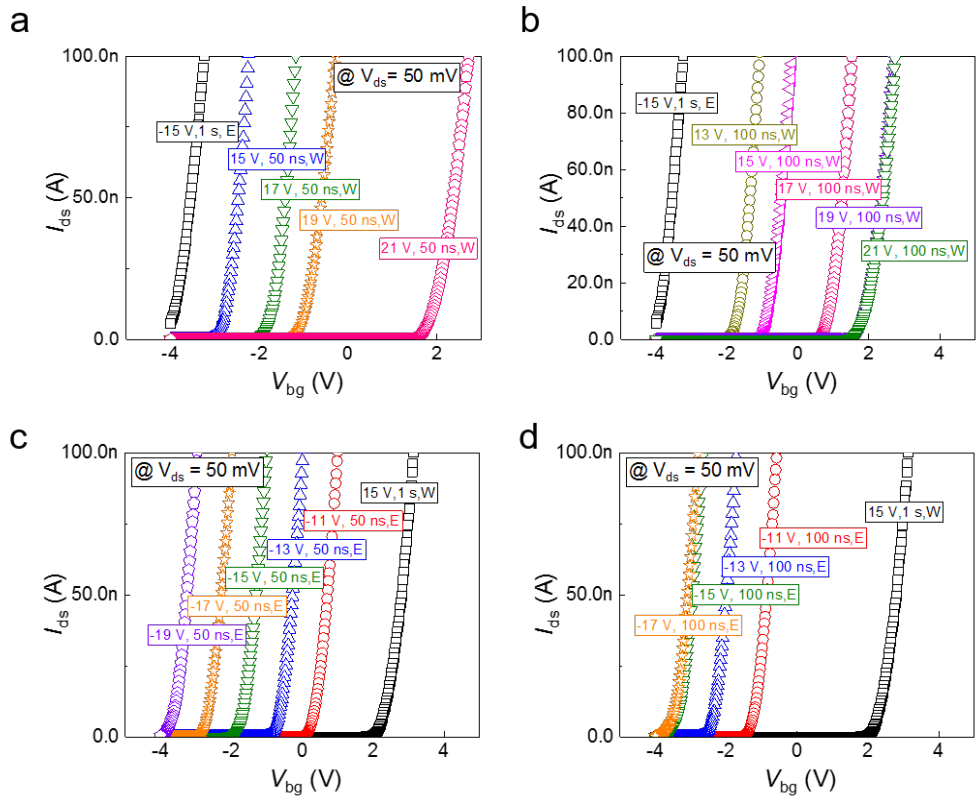

**Supplementary Figure 11**. Transfer characteristic curves of the 7.5 nm-thick h-BN flash memory device after different pulse operations.

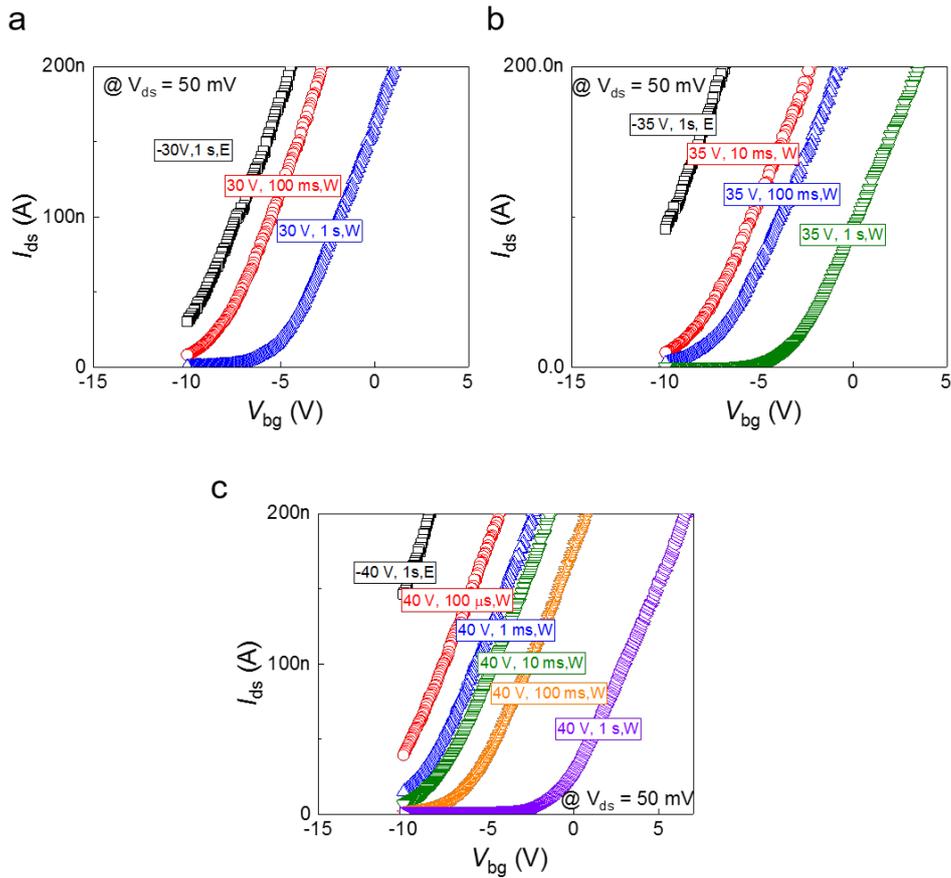

**Supplementary Figure 12.** Transfer characteristic curves of the 18 nm-thick h-BN flash memory device after different pulse operations.

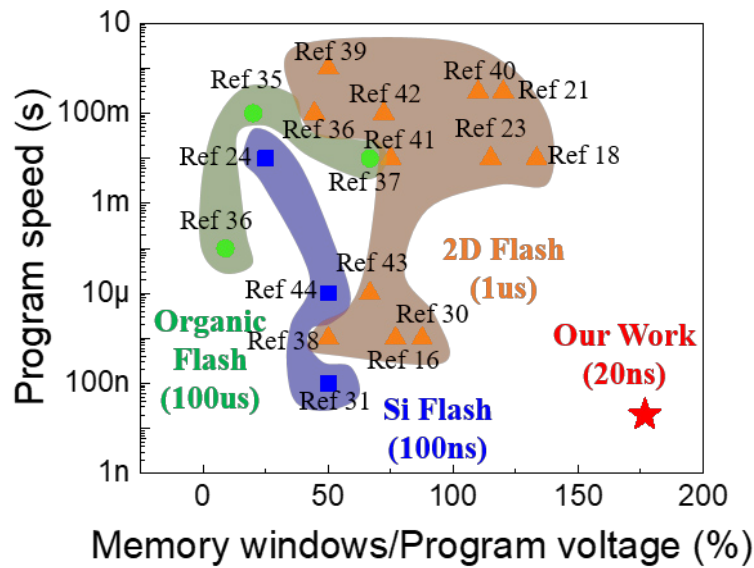

**Supplementary Figure 13.** Statistical comparison of programming speed and memory window/program voltage ratio of reported non-volatile flash memory, including organic flash

memory, 2D flash memory and silicon-based flash memory. For the first time, our flash refreshed the reported programming speed record of flash memory.